\documentclass[twocolumn]{emulateapj}
\usepackage{graphicx}
\usepackage{natbib}

\usepackage{float}
\usepackage{epsfig}
\usepackage{color}
\usepackage[colorlinks,urlcolor=blue,citecolor=blue,linkcolor=blue]{hyperref}

\begin{document}

\title{On Numerical Considerations for Modeling Reactive Astrophysical Shocks}

\author{Thomas L. Papatheodore\altaffilmark{1,2*}}
\author{O. E. Bronson Messer\altaffilmark{2,3,1$\dagger$}}
\affil{1. Department of Physics \& Astronomy, University of Tennessee, Knoxville, TN 37996}
\affil{2. National Center for Computational Sciences, Oak Ridge National Laboratory, Oak Ridge, TN 37831}
\affil{3. Physics Division, Oak Ridge National Laboratory, Oak Ridge, TN 37831}
\email{$^{*}$tpapathe@utk.edu, $^{\dagger}$bronson@ornl.gov}


\begin{abstract}
Simulating detonations in astrophysical environments is often complicated by numerical approximations to shock structure. A common prescription to ensure correct detonation speeds and associated quantities is to prohibit burning inside the numerically broadened shock \citep{fryxell1989}. We have performed a series of simulations to verify the efficacy of this approximation and to understand how resolution and dimensionality might affect its use. Our results show that, in one dimension, prohibiting burning in the shock is important wherever the carbon burning length is not resolved, in keeping with the results of \cite{fryxell1989}. In two dimensions, we find that the prohibition of shock burning effectively inhibits the development of cellular structure for all but the most highly-resolved cases. We discuss the possible impacts this outcome may have on sub-grid models and detonation propagation in models of Type Ia supernovae, including potential impacts on observables.
\end{abstract}

\section{Introduction}
Shocks are ubiquitous in astrophysical settings. In particular, the mechanisms which drive events such as novae, x-ray bursts, and Type Ia supernovae are all posited to involve detonations, a supersonic burning regime maintained by energy release from the propagation of a compressed, reactive shock. In all these cases, the explosive events are powered by rapid thermonuclear energy release and, therefore, share several characteristics. In this paper, we restrict our immediate attention to the most energetic of these thermonuclear events -- thermonuclear supernovae -- but the bulk of our results are generally applicable. 

Type Ia supernovae (SNe Ia) are defined spectroscopically by an absence of hydrogen, but the presence of strong silicon absorption lines near maximum light. Observations of these events show an inner region of iron-peak elements (iron, cobalt, nickel, etc.) with intermediate-mass elements (IMEs; elements between carbon and nickel) in the outer layers \citep{filippenko1997}, and it is the radioactive decay of newly synthesized nickel-56 ($^{56}$Ni $\rightarrow$ $^{56}$Co $\rightarrow$ $^{56}$Fe) that powers the optical light curves. SNe Ia are commonly attributed to thermonuclear explosions of carbon-oxygen white dwarfs (WDs) in binary stellar systems. The two main progenitor systems believed to give rise to these events are the single-degenerate (SD) scenario, where accretion of matter (hydrogen or helium) from a non-degenerate companion drives the WD toward the Chandrasekhar-mass until an explosion occurs, and the double-degenerate (DD) scenario, where an explosion results from the merger of a WD-WD binary pair. Each progenitor scenario includes various explosion mechanisms but there are only two modes of burning by which they can proceed: deflagration \citep{nomoto1976,gamezo2003} or detonation \citep{arnett1969}. 

In the most widely-studied version of the SD scenario, accretion continues until an explosion is triggered by carbon burning in the core of a near-Chandrasekhar-mass WD, but the exact details of the mechanism are unclear. For instance, once the thermonuclear runaway begins, does burning proceed as a deflagration or a detonation? Studies have been performed over the last several decades to answer such questions. Pure detonations have been shown to burn near-Chandrasekhar-mass WDs completely to iron-peak elements \citep{arnett1969,woosley1986} in contrast to observations, traditionally eliminating pure detonations as a suitable SNe Ia explosion mechanism. However, these studies were performed under hydrostatic conditions, whereas recent simulations indicate that consideration of rotating WDs \citep{pfannes2010} or sub-Chandrasekhar-mass WDs \citep{sim2010,vankerkwijk2010} may revive pure detonations in some cases. Deflagrations propagate sub-sonically, allowing time for expansion of the WD to lower densities where IMEs detected in spectra can be created. Although pure deflagrations have been shown to produce enough energy to explode a star \citep{hillebrandt2005}, they do not provide enough $^{56}$Ni to power the observed light curves, except perhaps in the lowest energy events. Another obstacle facing pure deflagration models is a substantial mixing of elements in the ejecta, in disagreement with observed distributions \citep{stehle2005}. Due to these contrary results, a widely-accepted mechanism for the SD channel is a deflagration that becomes a detonation at some point during the explosion. The main variations of this mechanism include the deflagration-to-detonation transition \citep[DDT;][]{khokhlov1991,gamezo2005,jackson2010}, pulsating-reverse detonation \citep[PRD;][]{bravo2006}, and gravitationally-confined detonation \citep[GCD;][]{plewa2004,jordan2008}. In these models a deflagration phase pre-expands the WD to lower density before a detonation occurs. At sufficiently low density, oxygen- and silicon-burning times can become comparable with the sound-crossing time of the WD, leading to incomplete burning and resulting in the production of IMEs. In the DDT model, the critical density ($\rho_{c}$) at which the transition to detonation occurs is an unknown parameter: Previous studies suggest \citep{gamezo2005} that $\rho_{c}$ \mbox{$\sim10^{7}$ g cm$^{-3}$} can produce results consistent with observations. Such parameterized models have been shown to reliably match light curves and elemental abundances \citep{kasen2009}, but further investigation is needed to help establish a theoretical understanding. 

In the DD scenario, a WD-WD binary pair spirals inward and eventually merges due to emission of gravitational waves, resulting in an explosion. Until recently, the DD scenario was considered by many to be an unreasonable explanation for SNe Ia because such a merger was thought to result in an accretion-induced collapse to a neutron star \citep{nomoto1991,maoz2012} instead of a thermonuclear supernova. However, modern simulations \citep{pakmor2010,pakmor2011,pakmor2012} show that explosions can occur if a detonation is ignited \textit{during} the merger process, and these explosions are consistent with the sub-luminous and normal classes of SNe Ia observations. The DD scenario has also received increasing support in recent literature due to merger rates consistent with SNe Ia rates \citep{wang2012}, as well as a natural explanation for the lack of hydrogen in the spectra. 

Based on energetics arguments, near-Chandrasekhar-mass progenitors have dominated the Type Ia literature. However, recent studies \citep{fink2010,sim2010,vankerkwijk2010,pakmor2012} indicate that sub-Chandrasekhar-mass systems might also be capable of explaining observations. Sub-Chandrasekhar-mass events may be described by the SD and/or DD scenarios, and they provide a simple explanation for the range of explosion energies, i.e. the (primary) WD mass. The SD branch of this model is described by the double-detonation mechanism \citep{nomoto1982a,nomoto1982b,woosley1994}, where a layer of helium is accumulated on the surface of a sub-Chandrasekhar-mass WD via accretion from a binary companion. A detonation may occur in the helium layer which drives shock waves into the core, inducing a subsequent carbon detonation at either the helium-carbon interface (edge lit) or the center of the WD (core compression). The DD branch has been postulated to result from a detonation during the violent merger of a WD-WD system with mass ratio near unity. This scenario can lead to a condensed object with mass less than the Chandrasekhar-mass which detonates in an envelope comprised of the remaining secondary WD material \citep{pakmor2010,pakmor2012}. The low-density environments common in sub-Chandrasekhar-mass WDs readily allow for the production of IMEs, eliminating the need for a pre-expansion phase. But as we shall see, simulating nuclear burning fronts under such low-density conditions requires caution, especially when complicated by multi-dimensional effects. In short, although there are competing hypotheses as to the progenitor system(s) of SNe Ia \citep[see][for further discussion of SNe Ia progenitors]{hillebrandt2000,nomoto2003,maoz2012,wang2012,nomoto2013}, almost all viable options involve a detonation at some point during the explosion. Therefore, accurate calculations of the propagation of detonation fronts are essential to making predictions which can be compared with SNe Ia observations. Calculations of adequate physical fidelity depend on a workable theory of detonations in astrophysical contexts and well-controlled computational implementations.

Detonation theory was first developed as a simple one-dimensional model by \citet{chapman1899} and \citet{jouguet1905} at the turn of the twentieth century. The Chapman-Jouguet theory (CJ-theory) models the detonation front as a sharp discontinuity between burned (ashes) and unburned (fuel) material. The reactions occur instantaneously as the interface propagates into the fuel, leaving completely burned ashes behind it. Given the energy released by burning fuel into ash, the detonation velocity and post-shock thermodynamic state can be obtained by this model but it fails to describe the structure of the detonation. The work of \citet{zeldovich1940}, \citet{vonneumann1942}, and \citet{doring1943} expanded CJ-theory by advancing reactions according to their corresponding rates, thereby describing a finite reaction zone with an extended thermodynamic profile behind the discontinuous shock (ZND-theory). In such one-dimensional models a \textit{burning length} can be defined as the product of the detonation velocity and the total time to achieve a particular ash state.

Modern hydrodynamics codes are capable of realistically modeling such stellar detonations, however, numerical approximations to shock structure can threaten the fidelity of such simulations if not properly treated. A \textit{physical shock} is considered to be infinitesimally thin under such conditions, so fuel spends very little time within the shock itself as the detonation propagates. This means that an insignificant amount of burning occurs naturally inside of a shock. The width of a \textit{numerical shock}, however, is constrained by the hydrodynamics scheme and spatial resolution employed in a simulation. For example, an Eulerian PPM scheme typically artificially widens a shock to approximately 2-4 computational zones, so on a sufficiently coarse grid an appreciable amount of burning may occur inside the shock (for other methods, the shock can be spread to 20 zones or more \citep[see][]{oran2000}). Any burning that takes place within the numerical shock does so under erroneous thermodynamic conditions since the density and temperature have not yet reached their post-shock values where the reactions would physically occur.

Astrophysical shocks are very often under-resolved because of the disparity of scales in these problems, and therefore the discrepancy between real and numerical shock structure can lead to behavior that is not realized in nature. Specifically, \cite{fryxell1989} showed that energy deposition in the leading edge of an under-resolved carbon-burning shock can generate a secondary shock structure, but is a numerical artifact, that propagates ahead of the true shock at speeds greater than the Chapman-Jouguet velocity. These results are unacceptable since both the increased detonation speed and double shock sequence disrupt the structure of the reaction zone. Behind the location of the true shock, the resulting abundances and thermodynamic state converge to the correct values when evolving the simulation with a conservative method, but conclusions which depend on specific aspects of the propagation and reaction zone details will be fundamentally flawed and possibly misinterpreted. A common prescription to ameliorate these nonphysical results is to prohibit burning inside of a numerical shock \citep{fryxell1989}, ensuring that burning does not occur until the correct post-shock conditions are reached. This approach eliminates the secondary shock structure and restores the proper detonation speed to the models. This remedy is especially relevant for astrophysical simulations where burning lengths are often smaller than the numerical shock width.

Despite the many successes of one-dimensional models, they do not agree with experimental results even when the vagaries of numerical versus physical shocks are taken into account. Real  (i.e. multi-dimensional) detonations propagate at a slightly reduced speed and exhibit a complex cellular pattern within the reaction zone. A physical burning front consists of alternating regions of Mach stems and incident shocks connected by transverse waves that extend back into the reaction zone (see section 2.2). The points where these three structures meet are called triple-shock configurations, or triple-points, and it is the paths of these high-pressure points which trace out the characteristic cellular pattern (see Section 2.2). Cellular detonations were first observed in terrestrial gases by \cite{denisov1959} and \cite{voitsekhovskii1963}. The cellular pattern can be recorded experimentally as the triple-points etch their paths on the inside of soot covered ``flame-tubes" \citep{Fickett1979}. Computational studies \citep{oran1998,gamezo1999a} show that cellular burning can create low-pressure pockets of unreacted gas, the presence of which increases the length of the reaction zone compared to one-dimensional results and can play an important role in the propagation and extinction of detonations. \cite{gamezo1999a} noted that the cell sizes and regularity of the final structure were only dependent on the chemical kinetics, and not on the peturbations used to disrupt the initially planar detonation front.

\cite{boisseau1996} were the first to demonstrate the existence of cellular detonations in degenerate carbon-oxygen fuel, representative of SNe Ia environments. They compared their results to the previous one-dimensional studies of \cite{khokhlov1989}, who calculated three distinct burning lengths ($x_{C} \ll x_{O} \ll x_{Si}$) corresponding to the burning time scales for carbon, oxygen, and silicon. Analogous to terrestrial detonations, \cite{boisseau1996} showed that cellular burning created pockets of incompletely burned material which altered their results relative to one-dimensional calculations. They found an increase in the length of the carbon-burning region, a slight decrease in detonation velocity, and changes in the resulting composition. Similar to \cite{gamezo1999a}, the structure that developed was shown to depend only on the material properties, and was independent of the initial perturbation, grid size, and boundary conditions. Although only the carbon-burning region was resolved in their study, \cite{boisseau1996} speculated on the importance of the silicon-burning cells, expressing the need to investigate their role in DDT and detonation quenching. 

Performing simulations of carbon-oxygen detonations at low SNe Ia densities \mbox{($1\times10^{6}-3\times10^{7}$ g cm$^{-3}$)}, \cite{gamezo1999b} investigated the different cell sizes corresponding to the one-dimensional \mbox{carbon-,} \mbox{oxygen-,} and silicon-burning regions. They found that cellular structure increased the length of the carbon- and oxygen-burning regions by a factor of 1.6 and the silicon-burning region by a factor of 1.3 relative to one-dimensional results. The increased burning lengths imply that incomplete burning can occur at higher densities than determined from one-dimensional calculations. \cite{gamezo1999b} explained that, considering these revised results, fits of delayed-detonation models to observation would yield a higher critical density at which DDT can occur. They estimated that $\rho_{c}$ could be increased by $\sim$6\% when the silicon-length is increased by $\sim$30\%. They also anticipated increased mixing of species due to the presence of incomplete and complete burning regions in the reaction zone, and suggested that this could be observable in the final composition distribution and velocities of the ejecta. The incomplete energy release from low-density burning was found to create conditions sufficient for detonation extinction in some cases.

\cite{timmes2000b} simulated cellular detonations in pure carbon to investigate how spatial resolution might affect such simulations. With a fuel density of \mbox{10$^{7}$ g cm$^{-3}$}, they showed that the size of the burning cells were not significantly affected by resolution, but the main features of the detonation front were strongly dependent. The curvature of the weak incident shocks and the strength of the triple-points and transverse waves were all shown to be influenced by the spatial resolution, and the differences between the incomplete and complete burning regions were also affected. These results help to define the minimum resolution necessary for multi-dimensional SNe Ia models involving cellular detonations.

In this work, we explore the effects of modifying the burning prescription in both one- and two-dimensional detonation simulations in order to fully understand the ramifications of this prescription, and, in particular, its effect on cellular burning. Simulations which intend to inform sub-grid flame models (for use in full-star simulations) with realistic detonation input must be capable of capturing the dynamic cellular burning front when densities of interest are such that detonation cellular sizes become comparable to scales just below the spatial resolution actually achieved in full-star simulations. As we shall see, an arbitrary prohibition on burning within a simulated shock can have a direct effect on the formation of cellular structure due to numerical considerations.

\section{Simulations}

\begin{figure}
\begin{center}
\includegraphics[scale=0.15]{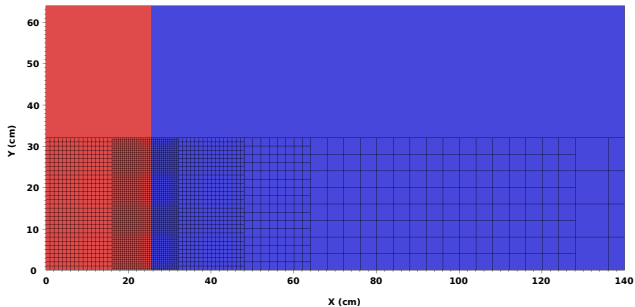}
\caption[center]{Temperature as a function of position for a small section of the problem setup along with the adaptive mesh (shown only on the bottom of the figure). Red \mbox{($1\times10^{10}$ K)} denotes the ``match head" region used to ignite the fuel shown in blue \mbox{($1\times10^{7}$ K).}} 
\label{fig:setup}
\end{center}
\end{figure}

This study was implemented using the multi-physics, parallel simulation code FLASH \citep[version 4;][]{fryxell2000}, which is capable of handling reactive compressible flow problems in astrophysical environments. The widely-used directionally-split piecewise-parabolic method \citep{colella1984} was chosen to evolve the Euler equations on an adaptive mesh, and these equations were closed using an equation of state relevant for electron-degenerate environments. Nuclear kinetics were advanced according to a 13-isotope $\alpha$-chain plus heavy-ion reaction network, Aprox13, which has been shown to capture energy generation rates for carbon and oxygen comparable to larger networks without their computational burden \citep{timmes2000a}. The reactions included ($\alpha, \gamma$) and ($\alpha, p)(p,\gamma$) links as well as their reverse sequences among the 13 isotopes $^{4}$He, $^{12}$C, $^{16}$O, $^{20}$Ne, $^{24}$Mg, $^{28}$Si, $^{32}$S, $^{36}$Ar, $^{40}$Ca, $^{44}$Ti, $^{48}$Cr, $^{52}$Fe, and $^{56}$Ni.

One- and two-dimensional detonations were modeled in a numerical laboratory which approximates conditions in the outer reactive regions of SNe Ia. The fuel consisted of a 50-50 carbon-oxygen composition at rest with a temperature of \mbox{$1\times10^{7}$ K}. Fuel densities were chosen separately for the one- and two-dimensional cases to demonstrate two distinct issues that can arise when modeling reactive shocks. The computational domain spanned 6144 cm in length by 128 cm wide with reflecting boundary conditions at the left (ash) end, outflow at the right (fuel) end, and periodic conditions along the length at $y=0$ and \mbox{$y=128$ cm}. This setup was chosen to resemble a small section of a detonating WD with adequate resolution to track the details of the reaction zone. The detonations were ignited by a ``match head" region with a temperature and x-velocity of \mbox{$1\times10^{10}$ K} and \mbox{$1\times10^{9}$ cm s$^{-1}$} at the left end of the domain, and then evolved to their self-sustained conditions \citep{sharpe1999} as they propagated down the ``tube". Figure \ref{fig:setup} illustrates the setup, including only a small section of the domain so that the match head region and the structure of the adaptive mesh may be discerned.

\subsection{One Dimension}
\begin{figure}[H]
\begin{center}
\includegraphics[scale=0.68]{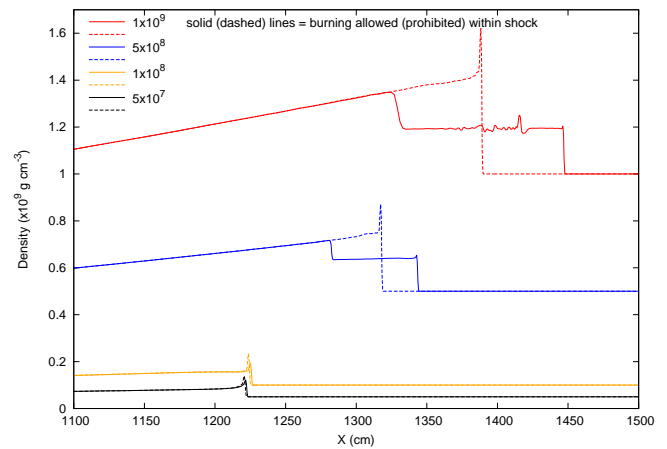}
\caption[center]{Results from the one-dimensional simulations showing the secondary shock structure in the density profile due to unresolved carbon-burning (0.5-cm resolution) at t $\sim$ \mbox{10$^{-6}$s}. The solid curves show the outcome when shock burning is allowed and the dashed curves show the results when burning is prohibited in the shock. The different pairs of curves correspond to the different initial fuel densities of $1\times10^{9}$, $5\times10^{8}$, $1\times10^{8}$, and \mbox{$5\times10^{7}$ g cm$^{-3}$}. We can see that preventing burning in the shock eliminates the nonphysical shock structure in the two higher-density simulations, leading to the correct detonation velocity and physically acceptable results.} 
\label{fig:1dCompare}
\end{center}
\end{figure}

\begin{figure}[t]
\begin{center}
\includegraphics[scale=0.68]{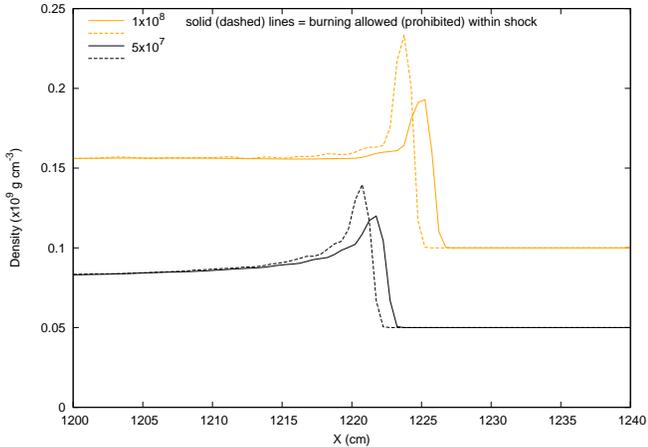}
\caption[center]{A zoomed in view of the density profiles from the $1\times10^{8}$ and \mbox{$5\times10^{7}$ g cm$^{-3}$} simulations. Although they do not show a secondary shock, there are differences in their peak values corresponding to premature burning conditions inside the shock as opposed to the proper post-shock conditions.}
\label{fig:1dCompareZoom}
\end{center}
\end{figure}

We simulated one-dimensional detonations with fuel densities of $1\times10^{9}$, $5\times10^{8}$, $1\times10^{8}$, and $5\times10^{7}$ \mbox{g cm$^{-3}$} in order to quantify and explain the issues found by \cite{fryxell1989} in under-resolved numerical shocks for a variety of SNe Ia regimes. These simulations were performed with a maximum spatial resolution of 0.5 cm, with burning allowed and then forbidden within the numerical shock. With no restrictions on burning, the $1\times10^{9}$ and \mbox{$5\times10^{8}$ g cm$^{-3}$} models were both found to exhibit the characteristic secondary shock wave described by \cite{fryxell1989} while the lower-density models did not. With burning prohibited in the shocks, the nonphysical shock waves were eliminated and correct detonation velocities were restored to the affected higher-density models, again in agreement with the results of \cite{fryxell1989}. This behavior is illustrated in Figure \ref{fig:1dCompare} which shows the resulting density profiles from the simulations. The solid and dashed curves in this figure correspond to the cases where burning within the shock is allowed and forbidden, respectively. Figure \ref{fig:1dCompareZoom} is a magnified view of these profiles, showing that although the two lower-density models do not exhibit a secondary shock, they differ in peak density since the case with burning within the shock forbidden processes the fuel at post-shock conditions while the case which allows burning in the shock does so under premature conditions. The observed offset of the front position between the two cases (with burning allowed and forbidden within the shock) is due to the numerical extent of the shock.

\begin{figure}[t]
\begin{center}
\includegraphics[scale=0.68]{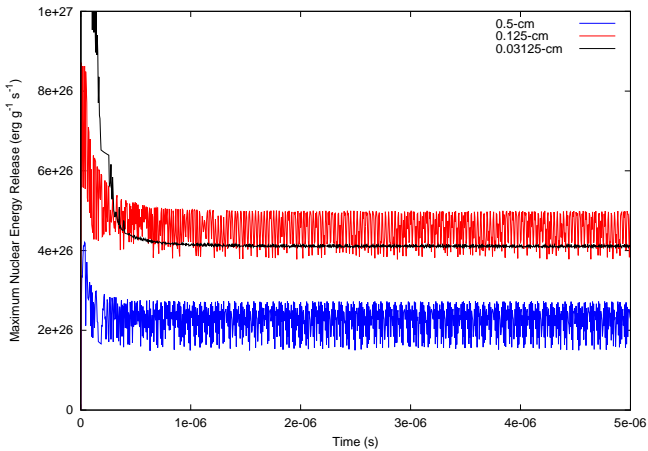}
\caption[center]{Maximum of nuclear energy release rate versus time for the $\rho_{i}$ = \mbox{$5\times10^{7}$ g cm$^{-3}$} simulation with 0.5-, 0.125-, and \mbox{0.03125-cm} maximum resolution. At 0.5-cm resolution the energy release is under-resolved and at \mbox{0.03125-cm} and above the rate has converged to its steady value.} 
\label{fig:enucVsTime}
\end{center}
\end{figure}

Significant adverse effects due to burning within the shock are only observed for the higher-density models since their carbon-burning lengths are smaller than those at lower densities. This allows a significant amount of carbon to be burned in the leading zones of the shock on a sufficiently coarse grid, creating a secondary shock structure. The speed of the nonphysical shock is greater than the CJ-velocity and is resolution-dependent \citep{leveque1998}, confirming it as a numerical artifact. Although the lower-density models are not afflicted by a secondary shock, it should be noted that all of these detonations are under-resolved. This can be seen in Figure \ref{fig:enucVsTime} which shows the maximum value of nuclear energy generation versus simulation time for $\rho_{i}$ = \mbox{$5\times10^{7}$ g cm$^{-3}$} at three different grid resolutions. Only the 0.03125-cm model is well enough resolved that the energy release has converged to the correct value. The oscillations in this figure are a numerical consequence of stepping through individual computational zones at the detonation front at discrete time intervals. 

\subsection{Two Dimensions}
We simulated two-dimensional detonations to determine how the treatment of burning in and around the numerical shock might affect cellular structure. First we performed a series of simulations without explictly perturbing the fuel, although numerical noise from roundoff errors can be sufficient to seed the transverse instability \citep{gamezo1999a} that leads to cellular burning. For comparison, we then performed a second series in which fuel perturbations were explictly defined. The perturbation was a 5-cm wide region, placed \mbox{20 cm} in front of the match head, in which the density of each zone was assigned a random value from a range 1\% above to 1\% below the density assigned to the rest of the flame tube. The detonations propagated in to fuel with a density of \mbox{$5\times10^{7}$ g cm$^{-3}$} and were simulated with burning allowed and forbidden within the shock as spatial resolution was improved. At this density, the grid was sufficiently refined to prevent the development of a secondary shock over the range of resolutions that we considered, thereby eliminating one-dimensional complications from the results. 

Here we have restricted our attention to densities greater than \mbox{$2\times10^{7}$ g cm$^{-3}$} (pathological regime) to avoid using over-driven detonations \citep{dominguez2011}. Indeed, all of our detonations begin with over-driven initial conditions but eventually relax to a steady, self-sustained state at which they propagate. At densities below \mbox{$2\times10^{7}$ g cm$^{-3}$}, our detonations decay and are eventually quenched as the over-driven support diminishes with time. It is possible to include constant support to these low-density detonations: \cite{gamezo1999b} effectively support their simulated detonations at low densities (\mbox{$1\times10^{6}$ - $3\times10^{7}$ g cm$^{-3}$}) by utilizing a moving grid. However, the structure of supported detonations differ from their unsupported counterparts \citep[higher temperatures lead to decreased burning lengths and change the subsequent energy release;][]{sharpe1999}. Therefore, we choose a density range to make connections with earlier works and to clearly delineate the effects of burning in numerical shocks.

\begin{figure*}[]
\begin{center}
\includegraphics[scale=0.25]{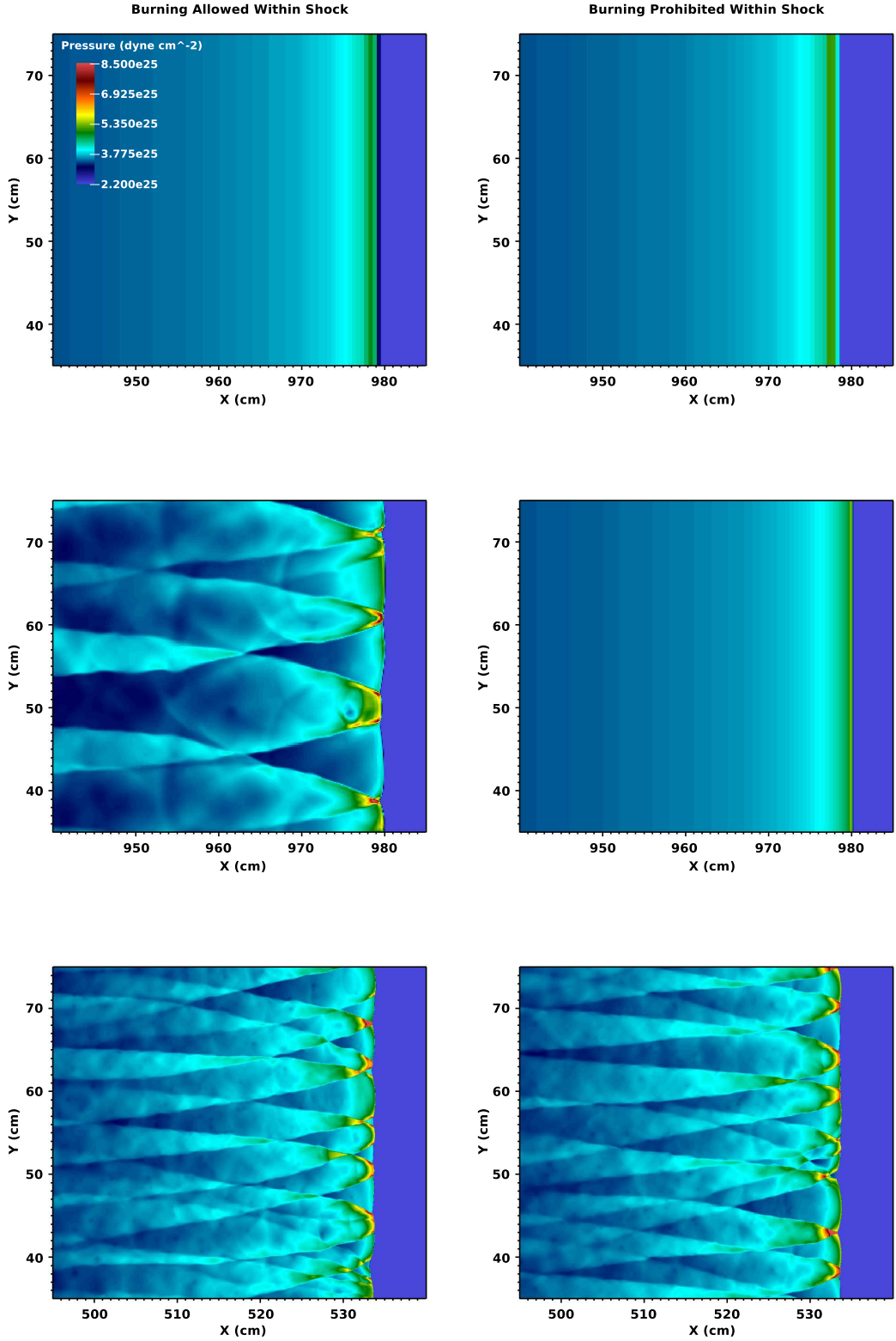}
\caption[center]{Snapshots of the resulting pressure from the unperturbed, two-dimensional series. The top, middle, and bottom rows show the results from the 0.5-, 0.125-, and 0.03125-cm maximum resolution simulations, respectively. The 0.5- and 0.125-cm panels show the results at t = \mbox{$8\times10^{-7}$s} and the 0.03125-cm panels show the results at t = \mbox{$4.15\times10^{-7}$s}. The differences in the fine features of the detonation fronts are due to various onset times for cellular burning.} 
\label{fig:comparePressure2D}
\end{center}
\end{figure*}

\begin{figure*}[]
\begin{center}
\includegraphics[scale=0.25]{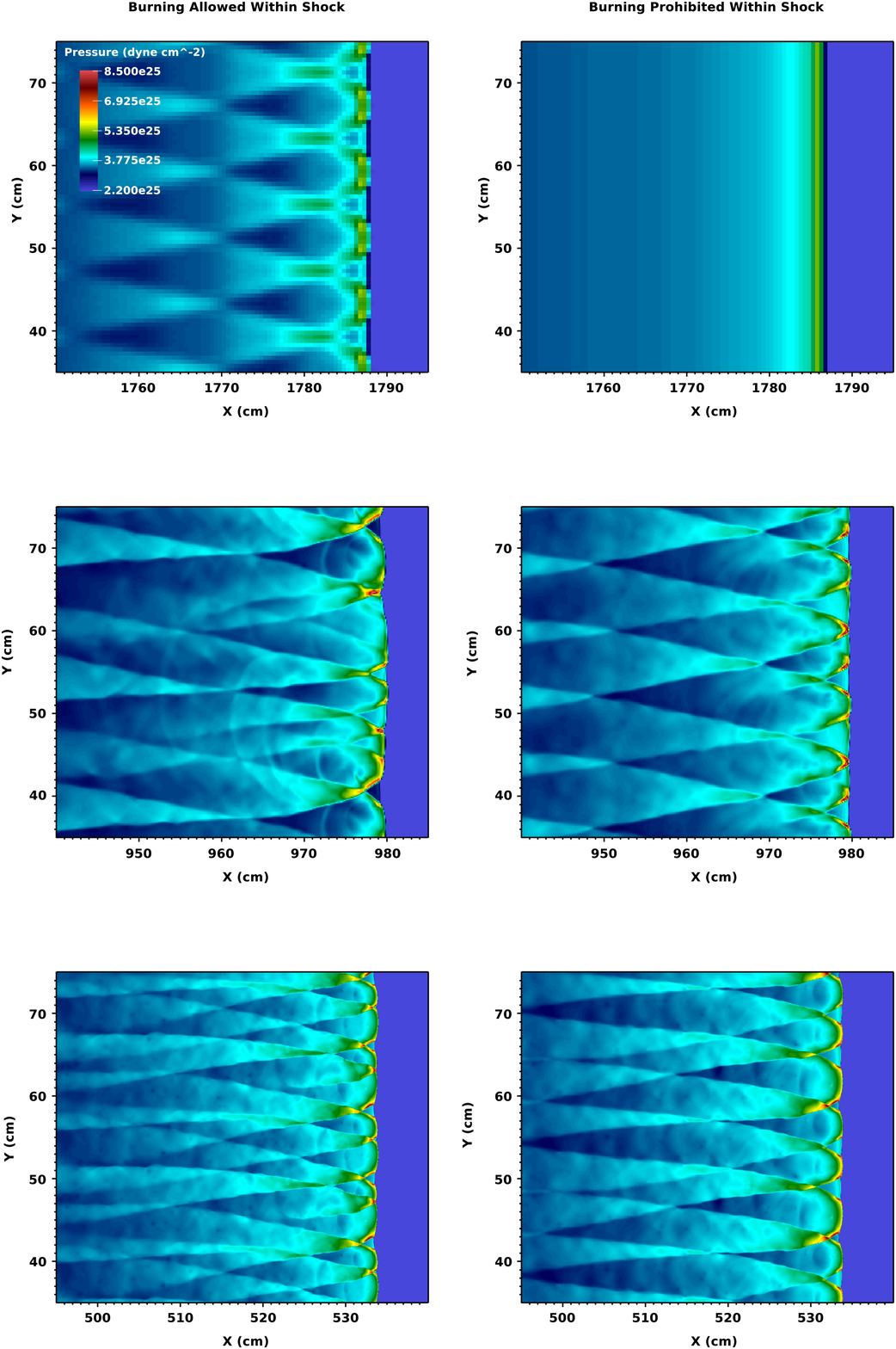}
\caption[center]{Snapshots of the resulting pressure from the perturbed, two-dimensional series. The top, middle, and bottom rows show the results from the 0.5-, 0.125-, and 0.03125-cm maximum resolution simulations, respectively. The 0.5-cm panels show the pressure at t = \mbox{$1.5\times10^{-6}$s}, the 0.125-cm panels show the results at t = \mbox{$8\times10^{-7}$s}, and the 0.03125-cm panels show the results at t = \mbox{$4.15\times10^{-7}$s}. The differences in the fine features of the detonation fronts are due to various onset times for cellular burning.} 
\label{fig:comparePressure2Dpert}
\end{center}
\end{figure*}

The results from the unperturbed series can be seen in Figure \ref{fig:comparePressure2D}. At the lowest level of resolution (0.5-cm) we observed that cellular structure was not established in either case, whether burning was allowed or forbidden within the shock. At 0.125-cm resolution the main difference between the two cases is apparent in that cellular structure developed where burning was allowed within the shock but did not develop where it was prohibited. At the highest resolution (0.03125-cm) cellular structure was observed in both cases. The results of the perturbed series were similar to those of the unperturbed series, except that only the 0.5-cm resolution simulations showed the discrepancy between the two cases, where burning was allowed and forbidden within the shock, while all higher-resolution runs exhibited cellular burning (See Figure \ref{fig:comparePressure2Dpert}).

\begin{figure}
\begin{center}
\includegraphics[scale=0.57]{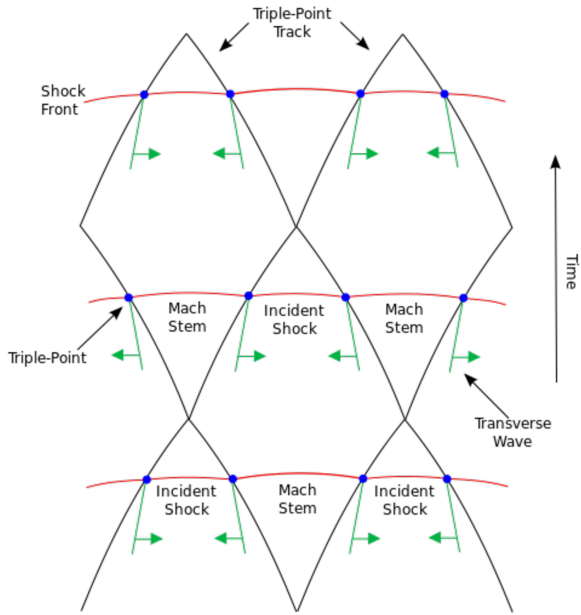}
\caption[center]{Diagram illustrating the formation of cellular structure. The shock front is shown at three times, superimposed on the triple-point paths \citep{Fickett1979,timmes2000b}. The main features of the detonation front are the Mach stems, incident shocks, and transverse waves. The points where these three features coexist are high-pressure regions called triple-shock structures, or triple-points, which trace out the cellular pattern shown in black.} 
\label{fig:cellular}
\end{center}
\end{figure}

In order to interpret these results, first consider the development of the transverse instability that leads to cellular burning. As an initially planar detonation propagates through a fuel source it inevitably encounters physical or numerical inhomogeneities within the fuel. These perturbations can disrupt the detonation front, introducing transverse motions to the fluid in the reactive flow behind the shock front. The magnitudes of these transverse motions grow as they are nurtured by nuclear energy release. Interactions at the detonation front between adjacent volumes of fluid with large, oppositely-oriented transverse velocities can create ``hot-spots" of increased density and temperature capable of igniting micro-detonations. These miniature detonations emit spherical shock waves which expand to overtake the main detonation front. These strong, expanding regions of the main front are called Mach stems, each of which is framed by a set of incident shocks. As neighboring Mach stems expand into the common incident shock lying between them, their transverse waves collide, creating a hot-spot which will then expand as a newly created Mach stem. Then, effectively, the segment of the detonation front that was an incident shock has evolved into a Mach stem and the segments of the front that were previously Mach stems are now the weaker incident shocks. The points on the front where a Mach stem, incident shock, and transverse wave coexist are called triple-shock configurations. These triple-points are high-pressure regions that move back and forth across the main front as the dynamic system propagates into the fuel, and it is these points that trace out the characteristic pattern of the cellular burning regime (see Figure \ref{fig:cellular}). Burning that occurs under the weaker conditions of the incident shock segments of the detonation front is incomplete relative to the burning that occurs at the Mach stem segments. This creates pockets of under-reacted, low-density material found by previous studies \citep{gamezo1999a,boisseau1996}.

Next consider a numerical interpretation of this scenario. A detonation is initially ignited with only an x-component of velocity and perturbations are introduced explicitly or through numerical noise on the grid. The perturbations induce the transverse fluid motions described above, however, in this case the shock is artificially stretched and dependent on grid resolution. On a coarse grid, if burning within the shock is prohibited, the burning and subsequent energy release are displaced away from their natural location just behind the detonation front. If the displacement is large enough, the transverse waves cannot interact properly with the detonation front and cellular burning will be inhibited.

Let us return to the recurrent picture of cell formation portrayed in Figure \ref{fig:cellular}, considering first the unperturbed series followed by the perturbed. In the unperturbed series, the 0.5-cm resolution simulations do not show cellular structure, regardless of their treatment of burning within the shock. This occurs because the energy release is not sufficiently localized to properly nurture the small transverse motions created by numerical noise. Figure \ref{fig:enucVsTime} shows the maximum rate of nuclear energy generation on the grid versus time, illustrating this inadequacy for the 0.5-cm resolution case. For the 0.125-cm resolution simulation where burning within the shock is allowed, cellular structure does occur. In this case the energy release is sufficient to develop the small transverse velocities to larger magnitudes. However, for the corresponding simulation where burning within the shock is prohibited, we do not see cellular structure. With the previous considerations in mind, we can see that this occurs because the energy deposition is displaced from the detonation front far enough that the resulting transverse components of the fluid fail to effectively interact with the front: The transverse velocities do not escalate and the dynamic triple-shock interactions cannot develop, inhibiting the formation of cellular structure. The 0.125-cm resolution simulations with burning allowed within the shock showed accumulated growth and interaction of the transverse components of the fluid until cellular structure developed. Conversely, for the case where burning within the shock was prevented, large, coherent regions of transverse velocities never developed. At the highest resolution (0.03125-cm), with burning prevented inside the shock, the transverse waves were able to interact with the front because the finer resolution leads to a numerical shock with smaller physical extent. Therefore, no qualitative differences are seen for the simulations at this resolution regardless of whether burning was allowed or prohibited within the shock. 

\begin{figure}[]
\begin{center}
\includegraphics[width=0.48\textwidth]{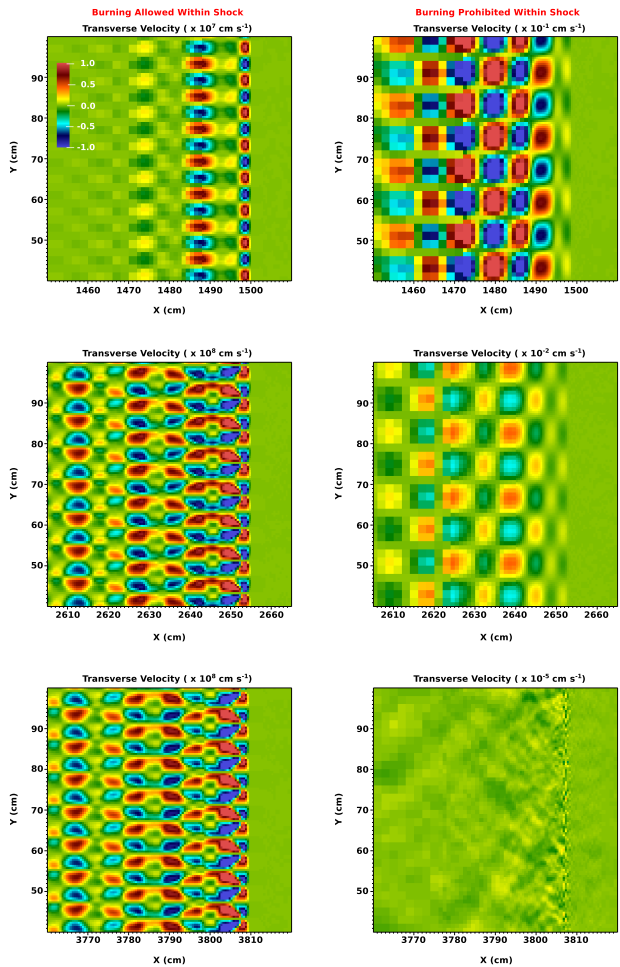}
\caption[center]{Transverse velocity versus position for the perturbed, 0.5-cm resolution model. The left and right panels show the cases where burning is allowed and prohibited within the shock, respectively. The color map can be seen in the top left panel only, while each panel has its own scale. The top, middle, and bottom rows show the results for the two cases at $1.25\times10^{-6}$ s, $2.25\times10^{-6}$ s, and $3.25\times10^{-6}$ s, respectively.}
\label{fig:transVel}
\end{center}
\end{figure}

Explicitly-defined perturbations imparted larger regions of higher-amplitude transverse motions to the fluid than those seeded by numerical noise, producing somewhat different results. Indeed, with explicit perturbations, the 0.5-cm resolution simulation with burning allowed within the shock exhibited cellular structure while its unperturbed counterpart did not. The energy release in the explicitly perturbed case did not need to nurture very small transverse motions to the point where they could effectively interact with the detonation front, as was required in the unperturbed series. Instead, the transverse motions, as created, were large enough that the energy release only needed to maintain the motion. In contrast, the 0.5-cm resolution model with burning prevented within the shock did not exhibit cellular structure in either the perturbed or unperturbed simulations. This is because the energy release was sufficiently displaced from the leading edge of the numerical shock that transverse motions could not be maintained near enough to the detonation front for the triple-point interactions to occur. This is illustrated in Figure \ref{fig:transVel}, which shows transverse velocities at the detonation front for the perturbed, 0.5-cm resolution simulations. The left column of panels shows a time sequence of the results when burning is allowed within the shock and the right column shows the same time sequence for the case when burning is prohibited within the shock. In the top panels both cases show structured transverse motions, owing to the perturbed initial conditions (although there is a marked difference in the scale of the magnitudes). However, as we follow the time sequence forward (down the columns), it is clear that when burning is allowed within the shock the transverse motions evolve to a stable structure but when burning is prevented within the shock the structure of the transverse motions decays until it is unrecognizable in the bottom panel. The increased spatial resolution found in the 0.125-cm resolution simulations with burning forbidden within the shock surmount this particular problem and cellular structure is seen in the perturbed series at this resolution, again due to the larger transverse motions imparted on the fluid. Because the numerical shock had smaller physical extent at the finer resolution, the stronger transverse waves were able to effectively interact with the detonation front from the time they were created. Conversely, the unperturbed simulations at this resolution require some time for the growth of very small transverse velocities before cellular structure can begin to form.

\begin{figure*}[t]
\begin{center}
\includegraphics[width=0.68\textwidth]{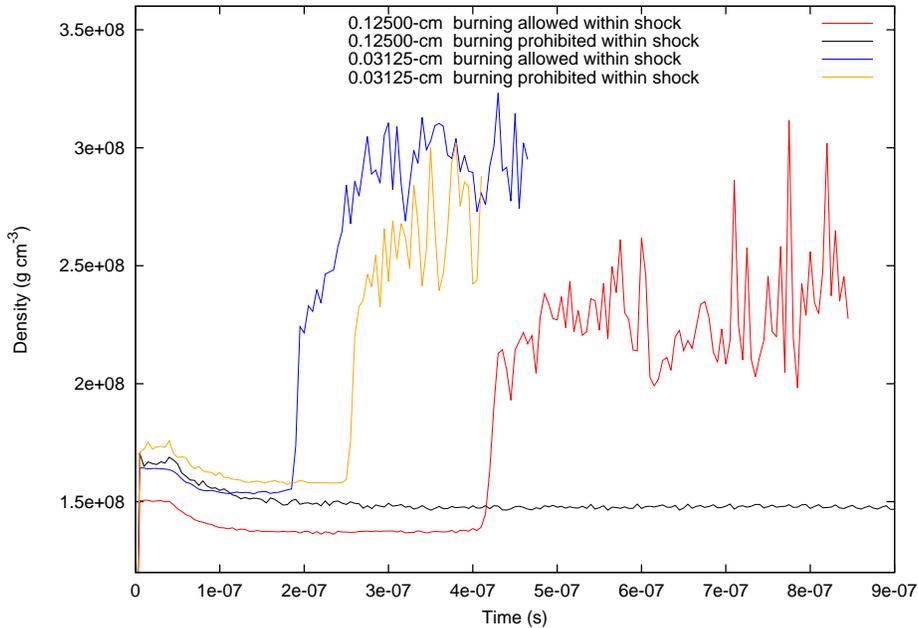}
\caption[center]{Maximum density profiles from the unperturbed, two-dimensional series showing the onset of cellular burning (steep increases in density). The strength of the triple-points corresponding to these maximum values are dependent on resolution as well as the treatment of burning within the numerical shock.} 
\label{fig:densityProfiles2D}
\end{center}
\end{figure*}

As in \cite{timmes2000b}, we find that the strength of the cellular features are resolution-dependent. Furthermore, we also find that these features are dependent on the treatment of burning within the shock. Figure \ref{fig:densityProfiles2D} is a plot of maximum density versus time for the unperturbed 0.125- and 0.03125-cm resolution simulations. For most of the evolution (i.e. always after the initial transient period and whenever cellular structure is achieved), these data correspond to the highest-density triple-point at each time step. Before the onset of cellular burning, it is the simulations that prevent burning within the shock which have higher densities, as burning does not occur until post-shock conditions are reached. However, during cellular burning, it is the simulations that allow burning within the shock which exhibit the highest densities. With burning allowed within the shock the triple-points are enhanced by the multi-dimensional dynamic effects of energy release within the leading zones of the main shock front. This ``pre-conditions" the fluid before the triple-points are fully formed, increasing the density and temperature within the triple-points. However, the corresponding consumption of fuel within the shock leaves less fuel available to burn in the triple-points. These effects conspire to result in lower values of instantaneous maximum energy release in the triple-points than was the case where burning was prohibited within the shock.

\section{Summary and Discussion}
We find that prohibition of burning within a numerically-widened shock is important when a significant amount of fuel will burn within it, consistent with the results of \cite{fryxell1989}. A sufficiently resolved shock does not require application of this prescription, however its promiscuous use can ensure integrity without demanding a search for adequate refinement at different fuel densities. Hence the well-known recommendation to prevent burning inside of a simulated shock. In two-dimensions we find that imposing this treatment can unintentionally inhibit the formation of cellular structure by displacing the release of nuclear energy to the region behind a coarsely resolved shock, therefore obstructing the development of transverse perturbations. In addition, we find that the strength of the triple-points, as well as the energy release realized in the triple points, is affected by the treatment of shock burning. 

Due to the disparity of length scales involved in SNe Ia, sub-grid models are required to approximate burning in full-star explosion models, and it is direct numerical simulations (DNS) such as those presented here that are used to inform these sub-grid models. We assert that our results may influence the fidelity of large-scale SNe Ia simulations through this channel. Simulations which model multi-dimensional detonation fronts in Type Ia environments \citep{boisseau1996,gamezo1999b,timmes2000b} have shown that pockets of incompletely-burned material, created by the presence of cellular structure, can increase the effective size of burning regions relative to one-dimensional calculations. Because the production of IMEs would be possible at higher densities, \cite{gamezo1999b} argued that this increases the critical density at which a DDT can occur. Our results suggest this effect can be hidden due to a particular treatment of burning within the numerical shock, either in a sub-grid model or \textit{in situ} at very low density. Importantly, these effects are most prominent in sub-Chandrasekhar-mass models due to their larger amounts of material at low densities. Recent simulations \citep{fink2010,sim2010,vankerkwijk2010,pakmor2012} have shown that pure detonations of sub-Chandrasekhar-mass WDs commonly result in incomplete burning due to these low densites \mbox{($<$ $10^{7}$ g cm$^{-3}$)}. The elongation of burning lengths due to cellular structure can exacerbate the incompleteness of burning in these cases: At a given time, cellular formation will lead to more partially burned material behind the shock. For unresolved cellular formation, a sub-grid model that does not account for this modification will lead to incorrect energy release and isotopic composition. At lower densities, cellular structure might well be resolved (albeit coarsely) in large-scale simulations. At these densities, the prohibition of burning within the shock will effectively retard this effect, leading again to incorrect results.

Detailed abundances are required for use in comparison to observations, especially at low densities \mbox{($<$ 10$^{7}$ g cm$^{-3}$)}. A sub-grid model will be necessary to account for multi-dimensional, small-scale structure if this comparison is to be of predictive value. Although previously thought to be rare in Type Ia spectra \citep[believed to be mostly related to super-Chandrasekhar-mass events, e.g.][]{howell2006}, recent observations \citep{parrent2011,silverman2012,zheng2013} have shown the presence of unburned carbon (\mbox{C$_{II}$ $\lambda$6580} features) in the early spectra of a significant number \mbox{($\sim$ 30\%)} of all events. It may be that these features are possible in sub-Chandrasekhar-mass \textit{and} near-Chandrasekhar-mass channels as well. Pockets of unburned carbon formed by cellular burning may remain if nuclear burning is arrested due to expansion in low-density material, essentially ``freezing out" carbon in the outer ejecta. Failure to account for cellular burning would hide the presence of unburned carbon from this process. In the case of near-Chandrasekhar-mass models, the amount of material at these densities is small, but in sub-Chandrasekhar-mass models there are significant amounts of mass at these densities. The ability to discriminate between the two models depends on a quantatative calculation of the total amount of unburned carbon. In order to determine if ``freezing out" of unburned carbon pockets actually impacts carbon abundance predictions from simulations, care must be taken to obviate the effects of burning within the numerical shock.

\acknowledgements
This work was sponsored in part by the Laboratory Directed Research and Development Program (Seed Money Fund) of Oak Ridge National Laboratory, managed by UT-Battelle, LLC, for the US Department of Energy. This research used resources of the Oak Ridge Leadership Computing Facility at the Oak Ridge National Laboratory, which is supported by the Office of Science of the U.S. Department of Energy under Contract No.~DE-AC05-00OR22725. The authors thank Raph Hix, Suzanne Parete-Koon, Chris Smith, and Dean Townsley for useful discussions. The authors also thank the anonymous referee for suggestions that materially improved this paper.


\end{document}